\documentclass[twocolumn]{aastex62}

\accepted{May 7, 2018}

\submitjournal{ApJ}

\shorttitle{BAaDE: SiO masers}
\shortauthors{Trapp et al.}

\begin{document}

\title{SiO Masers in the Galactic Bulge and Disk: Kinematics from the BAaDE Survey}

\author{A.C. Trapp}
\affil{Department of Physics and Astronomy, University of California, Los Angeles, CA 90095-1562, USA}

\author{R.M. Rich}
\affil{Department of Physics and Astronomy, University of California, Los Angeles, CA 90095-1562, USA}

\author{M.R. Morris}
\affil{Department of Physics and Astronomy, University of California, Los Angeles, CA 90095-1562, USA}

\author{L.O. Sjouwerman}
\affil{National Radio Astronomy Observatory, 1003 Lopezville Rd, Socorro, NM 87801}

\author{Y.M. Pihlstr{\"o}m}
\affil{Department of Physics and Astronomy, University of New Mexico, Albuquerque, NM 87131, USA}
\affil{National Radio Astronomy Observatory, 1003 Lopezville Rd, Socorro, NM 87801}

\author{M. Claussen}
\affil{National Radio Astronomy Observatory, 1003 Lopezville Rd, Socorro, NM 87801}

\author{M.C. Stroh}
\affil{Department of Physics and Astronomy, University of New Mexico, Albuquerque, NM 87131, USA}





\begin{abstract}
We present the first results from the BAaDE (Bulge Asymmetries and Dynamic Evolution) survey. Though only a subset of the complete survey ($\sim$2700 out of $\sim$20000 final sources), our data comprise the largest radio kinematic survey to date of stellar SiO masers observed toward the Galactic bulge and plane from $-15^\circ < l < +12^\circ$ and $-6^\circ < b < +6^\circ$. Our sources include a substantial number of line-of-sight (LoS) velocities in high extinction regions within $\pm 1^\circ$ of the Galactic plane. When matched with 2MASS\footnote{The Two Micron All Sky Survey, Skrutskie et al. (2006)} photometry, our radio-detected sample lies significantly brighter than and red-ward of the first red giant branch tip, reaching extremes of $(J-Ks)_0 > 8$, colors consistent with Mira variables and mass losing AGB stars. We see a clean division into two kinematic populations: a kinematically cold ($\sigma \sim 50\rm\ km\ sec^{-1}$) population that we propose is in the foreground disk, consisting of giants with 2MASS $Ks<5.5$, and a kinematically hot ($\sigma \sim 100\rm\ km\ sec^{-1}$) candidate bulge/bar population for most giants with $Ks>5.5$.  Only the kinematically hot giants with $Ks>5.5$ include the reddest stars. Adopting 8.3 kpc to the Galactic Center, and correcting for foreground extinction, we find that most of the sources have $M_{bol} \sim -5$, consistent with their being luminous, and possibly intermediate age, AGB stars. We note some tension between the possibly intermediate age of the kinematically hot population, and its high velocity dispersion compared to the disk. 
\end{abstract}





\section{Introduction} \label{sec:intro}

Some spiral galaxies are barred, and some are not. It is of fundamental importance to the study of galaxy formation to understand why this is the case. Bars can dominate the stellar fraction of the galactic bulges they are in (e.g. the Milky Way; Rich et al. 2007); can create torques leading to radial transport of interstellar gas that may fuel AGN; and are stable enough to last billions of years, yet simply do not form in a large fraction of otherwise normal spiral galaxies (Sellwood 2014).

Understanding of the formation conditions of bars requires detailed knowledge of their current structure, and the Milky Way hosts the only bar/bulge for which 3-D kinematics can be measured for the foreseeable future. The intensity of various efforts to survey the Milky Way bulge is a reflection of this. The BRAVA survey is a radial velocity kinematic study of the bulge with large numbers of stars (Rich et al. 2007), the ARGOS survey extended stellar kinematic surveys to greater Galactic longitude (Ness et al. 2013a, 2013b), and the GIBS survey pushes to lower Galactic latitudes (Zoccali et al. 2014, 2017). The use of infrared techniques to push infrared spectroscopy deep into the Galactic plane has just begun with APOGEE and VVV and has even reached the red clump (Babusiaux et al. 2014).

Extreme dust extinction is one of the defining hurdles in all of the above surveys, even at 2$\mu m$. Our program uses red giant maser sources emitting at radio wavelengths that are unattenuated by extinction and offer one of the best opportunities to pierce the dusty veil of the Galactic plane. The combination of an intense infrared radiation field re-emitted from dust in the stellar envelope and a high abundance of molecules in the cool stellar atmosphere provides the conditions needed for the level inversions that give rise to maser emission (see e.g. Habing 1996).  While masers are mostly found in the expanding circumstellar shells of AGB stars, they can also occur in normal red giants; many maser sources are matched with long-period Mira variables (see Messineo et al. 2002 for SiO masers).  Two fundamentally different types of masers can be used to explore the AGB population: OH masers, which arise in the expanding shell $\sim$1000 AU from the core, are found in the most evolved AGB stars, and are relatively rare (see, e.g., Habing 2016), and SiO masers, which arise only $\sim$10 AU from the core and are also associated with luminous mass-losing AGB stars, but are far more common.  In the case of the SiO masers, the radio emission lines give precise LoS (line-of-sight) velocities of their host star to $\sim$5$ \rm~km~s^{-1}$.  

Here we report on early results from the Bulge Asymmetry and Dynamical Evolution (BAaDE) experiment, which has been described in Sjouwerman et al. 2015, 2017: a survey of SiO maser sources near the Galactic plane. The program, techniques, and additional early results are fully described in Sjouwerman et al. (2018) (in prep.), which will also publish the catalog of sources we report on here.  This program succeeds in detecting tens of thousands of SiO masers by employing improved mid-infrared color-based candidate selection from the MSX\footnote{Midcourse Science Experiment, Price et al. (2001)} survey (Sjouwerman et al. 2009) and the sensitivity of the VLA\footnote{The Karl G. Jansky Very Large Array} and ALMA\footnote{Atacama Large Millimeter/submillimeter Array}.  However, the key has been a new technique that permits a rapid survey of candidates without the need to separately calibrate each detected source. BAaDE aims to explore the kinematics of this population and to extend studies of the bar and plane into regions of high stellar density and extinction that are challenging for optical and near-IR spectroscopy.  As the masers are signposts for AGB stars, SiO maser sources can be the progeny of intermediate-age stellar populations and may make it possible to explore the dynamics and origins of stellar populations in the 2-8 Gyr range. 
While a more detailed consideration of age is outside of the scope of this paper, there are indications from Mira periods and luminosities that at least some of our maser sources are younger than 10 Gyr.

\begin{figure*}
   \centering
   \includegraphics[width=\hsize]{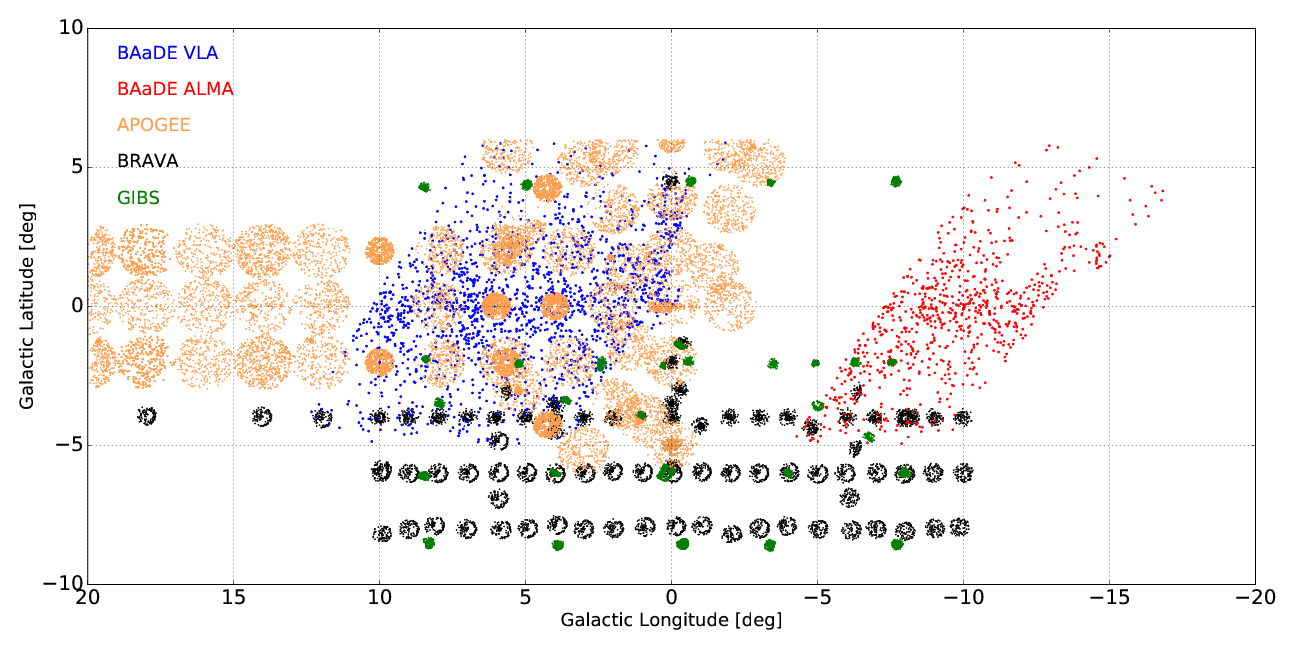}
   \caption{Positions of individual sources for major bulge optical/IR surveys with public datasets, including our BAaDE survey, BRAVA radial velocity survey (Rich et al. 2007), APOGEE near-IR survey (Nidever et al. 2012b), and GIBS (Zoccali et al. 2014). The legend gives the colors corresponding to the various surveys.}
   \label{POS}%
\end{figure*}

\begin{figure*}
   \centering
   \includegraphics[width = \hsize]{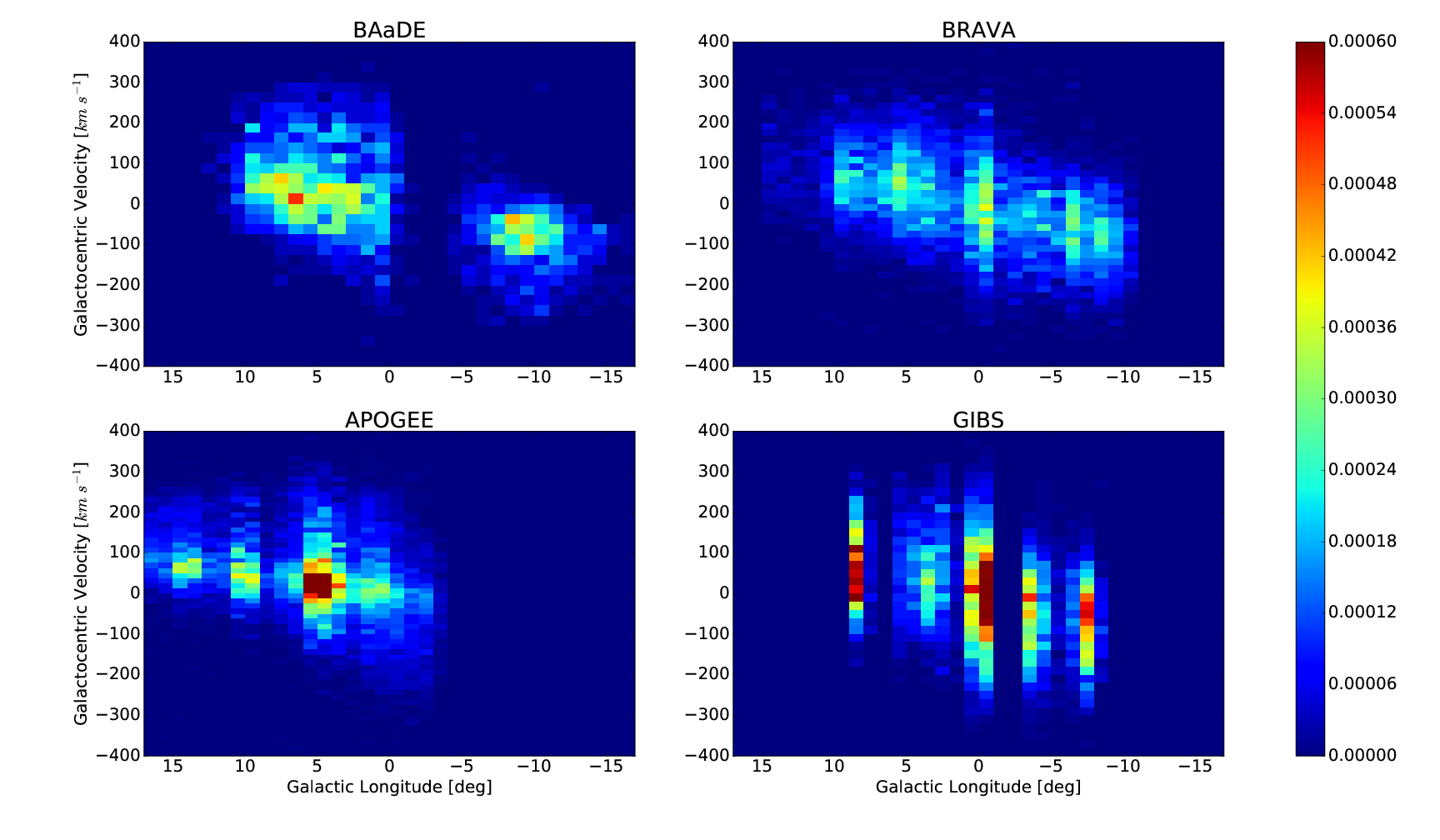}
   \caption{Longitude-LoS velocity 2D histograms (normalized) illustrate the kinematics of the BAaDE survey, compared to the three other major bulge surveys we are considering. To first order, the velocity dispersion and rotation of BAaDE are in agreement with the other studies.  Note that while all four studies survey the bulge, their sample selection differs.}
   \label{ALL_KIN}%
\end{figure*}

\begin{figure}
   \centering
   \includegraphics[width=\hsize]{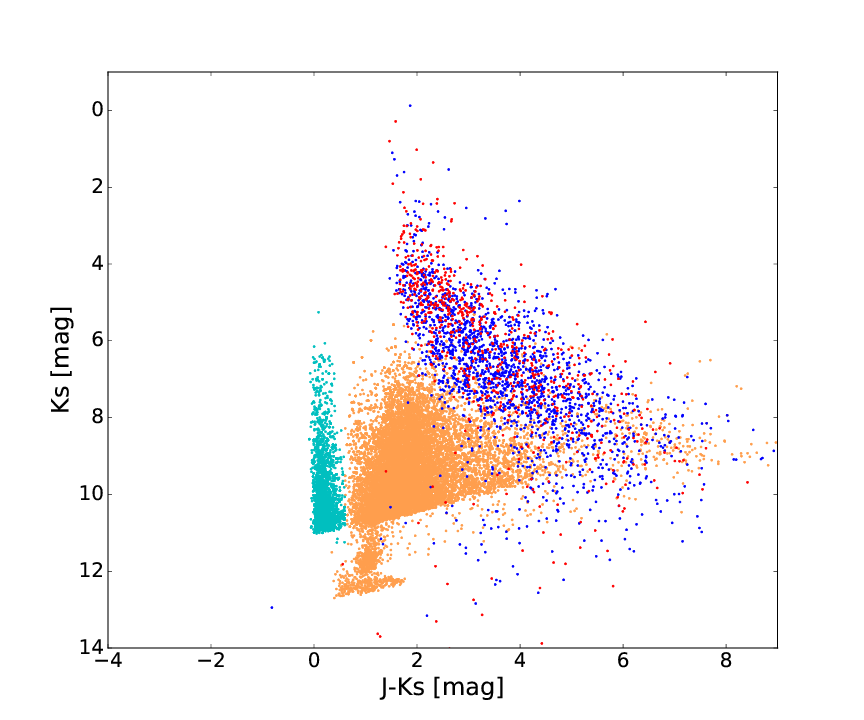}
   \caption{2MASS CMD of source identifications for BAaDE VLA and ALMA (blue and red) and APOGEE giants and main sequence stars (orange and cyan). These samples have almost no overlap in color-magnitude space (see section \ref{sec:pop_analysis}).}
   \label{CMD_1}%
\end{figure}

The appreciation that our Galaxy is barred emerged initially from modeling the dynamics of gas (Liszt and Burton 1980) with convincing evidence following from  the models of the bar's appearance in projection, as seen in infrared light (Blitz and Spergel 1991a, 1991b; Dwek et al. 1995).   The use of red clump stars as approximate distance indicators convincingly demonstrated the bar's angle and depth (Stanek et al. 1997; Babusiaux and Gilmore 2005a, 2005b).  Finally, observed kinematics of the giants showed the bar to be the predominant mass with little allowance for a classical bulge (e.g. the BRAVA survey; Rich et al. 2007; Shen et al. 2010; Kunder et al. 2012).  Indeed, the bar as the dominant population, including the X-shaped subpopulation, has been established via photometric models (Wegg \& Gerhard 2013), although recent work (Simion et al. 2017) has raised questions about the mass fraction of stars in the X-shaped bulge. 

The age of stellar populations in the bar/bulge is a current topic of debate.
Catchpole et al. (2016) report a survey of Mira variables that show their positions are a function of age. While the long period Miras ($P>400$ day) appear to trace the alignment of the bar major axis, the short period Miras ($P<400 $ day) do not. Recalling that the longest known period for a Mira in a globular cluster is 310 days (Feast et al. 2002), one may reasonably associate the short-period Miras with the oldest stellar populations, and the long-period Miras -- and by extension, the bar -- with intermediate age.  It has been shown that the bar accounts for most of the dynamical mass in the bulge (Shen et al. 2010; Portail et al. 2015); the Mira results would imply this mass to be $<10$ Gyr old.

The only {\it demonstrated} $>10$ Gyr old stellar population in the central region are the RR Lyrae stars, which have far lower rotation and higher velocity dispersion than the bar (Kunder et al. 2016).   When a direct measurement of age is not available, the kinematics can provide interesting constraints; We will use the kinematics of the SiO masers to place them, as best we can, among the hierarchy of bulge populations.

We define an intermediate age stellar population here as any population that is demonstrably younger than globular clusters and RR Lyrae stars.  The age range for ``intermediate" is typically 2-10 Gyr.
An argument has been made for intermediate-age stars in the bulge based on high-resolution spectroscopy of 99 bulge dwarfs strongly amplified by micro-lensing (Bensby et al.\ 2017).  Employing a self-consistent spectroscopic analysis to derive [Fe/H], $log \ g$, and $T_{eff}$, Bensby et al.\ argue that a substantial fraction of stars with [Fe/H] $> -0.5$ must be placed on intermediate-age isochrones. If such a stellar population is present in the bulge, stellar evolution requires that its progeny include many luminous AGB stars, and perhaps the longer-period Miras of Catchpole et al. (2016) represents that population.  Ness et al. (2013a) argue that the metal-rich subpopulation of the bulge is preferentially concentrated toward the plane, and Ness et al. (2014) argue that in formation scenarios with a disk origin, such young populations are expected. Taken at face value, the evidence would appear to powerfully argue for the bar and its populations being young or intermediate age. Yet the predominance of $\sim 10 $ Gyr old stars, and absence of young stars above the turnoff in the HST proper-motion-separated color-magnitude diagrams of Kuijken \& Rich (2002) and Clarkson et al. (2008a, 2008b, 2011) pose a problem.  
A recent analysis using the proper-motion cleaned HST Bulge Treasury dataset argues strongly for an ancient strongly peaked starburst (Gennaro et al. 2015). However, Haywood et al. (2016) propose an alternative model of the color-magnitude diagram arguing for a young population contribution. Bernard et al. (2018) use the HST proper-motion separated CMDs, and find that $~11\%$ of the bulge is younger than 5 Gyr. The complexity of modeling a faint, foreground contaminated main sequence turnoff population will likely leave this question unsettled, but it is clear that the HST photometry is not favoring a prominent, easily detected intermediate age population.

Groundbased direct study of the bulge turnoff began with Terndrup (1988), but modern studies employ foreground disk population subtraction, beginning with Ortolani et al. (1996), and following with Zoccali et al. (2003), Feltzing and Gilmore (2004) and Valenti et al. (2013), all of whom find predominantly 10 Gyr old populations from main-sequence turnoff photometry.  If a population of 1-5 Gyr old main-sequence stars is present in the bulge, it is proving very difficult to isolate directly. Although the Miras of period $>$ 400 days are nominally associated with intermediate-age stars, the high metallicity in the bulge may allow low-mass stars to attain an unexpectedly high luminosity as AGB stars (see e.g. Guarnieri et al. 1996).  Although we do not consider the age of the bulge in this study, we will conclude that the SiO maser population is more likely to be associated with the long-period Miras and that it is therefore a nominally intermediate-age population.

Other studies have employed SiO masers as kinematic probes, although ours has the largest numbers. The pioneering work done with the Nobeyama 45-m telescope in Japan (e.g., Izumiura et al. 1994, Deguchi et al. 2000a, Deguchi et al. 2004b, Fujii et al. 2006) detected hundreds of SiO maser sources across the bulge and found evidence of rotation. These early investigations generally used the IRAS satellite surveys and surveys of Miras to select targets, finding that the masers have red IR colors and a rising detection rate as a function of Mira period (Imai et al. 2002).  A master catalog of the Nobeyama observations is in preparation (Nakashima 2018, Private Communication) and we will compare our observations to that sample in future work. There have also been extensive bulge kinematic studies using OH/IR masers with numbers in the hundreds (e.g. Lindqvist et al. 1992; Sevenster et al. 1995, 1997; Sjouwerman 1998). These works, as well as others (e.g. Messineo et al. 2002), will provide invaluable comparisons and building blocks for this next-generation radio-measured kinematic bulge survey.  

Our full sample will contain LoS velocities throughout the entire bulge. Previous large optical/near-IR bulge surveys have surveyed areas of lower extinction, farther from the plane of the Galaxy, or on the near side of the bulge. We compare our data to three of these major bulge surveys that have made their data public: the Bulge Radial Velocity Assay (BRAVA, Rich et al. 2007 and Kunder et al. 2012), the GIRAFFE Inner Bulge Survey (GIBS, Zoccali et al. 2014), and the Apache Point Observatory Galactic Evolution Experiment (APOGEE, Nidever et al. 2012b). BRAVA and GIBS probed latitudes $2^\circ < |b| < 8^\circ$, and longitudes $-10^\circ < l < +10^\circ$. APOGEE reaches to the Galactic plane, and measures abundances and individual extinctions for much of their sample, but does not cover longitudes $l < -3^\circ $. Although Babusiaux et al. (2014) surveys red clump giants in the plane in specific fields, BAaDE is the first survey that more or less continuously spans a range in Galactic longitude over $-2< b^\circ <+2$.  See Fig.~\ref{POS} for spatial coverage of the various bulge surveys and Fig.~\ref{ALL_KIN} for their kinematic comparisons; we have not included that of Babusiaux et al. (2014) because it sparsely samples the region of the BAaDE survey.

The sample presented here consists of the first subsample of the BAaDE survey, and allows the measurement of line of sight velocity distributions (LOSVDs) close to the Galactic plane ($|b| < 6^\circ $) with longitudes $-15^\circ  < l < 12^\circ $. With our current data and thousands more LoS velocity measurements underway, we hope to constrain dynamical models of the Milky Way Bulge. Observations have been made that will fill in the gap between the ALMA and VLA observations in Fig.~\ref{POS} ($-5^\circ <l<0^\circ $), allowing for a continuous measurement of the rotation curve and velocity dispersion.

In section \ref{sec:data}, we present a summary of the data acquisition and analysis. In section \ref{sec:pop_sep}, we separate our stars into three kinematic sub-populations using the 2MASS\footnote{The Two Micron All Sky Survey, Skrutskie et al. (2006)} color-magnitude diagram and our LoS velocities. Section \ref{sec:pop_analysis} analyzes the rotation curves and other properties for the different subgroups, allowing us to assign them to either the disk or bulge population. We present a new analysis of the LOSVD in section \ref{sec:skew}, and a candidate new kinematic component near $-200\rm~km~s^{-1}$ in section \ref{sec:200KMS}, and finally, we give a summary of results in the conclusion section.

\begin{figure*}
   \centering
   \includegraphics[width=\hsize]{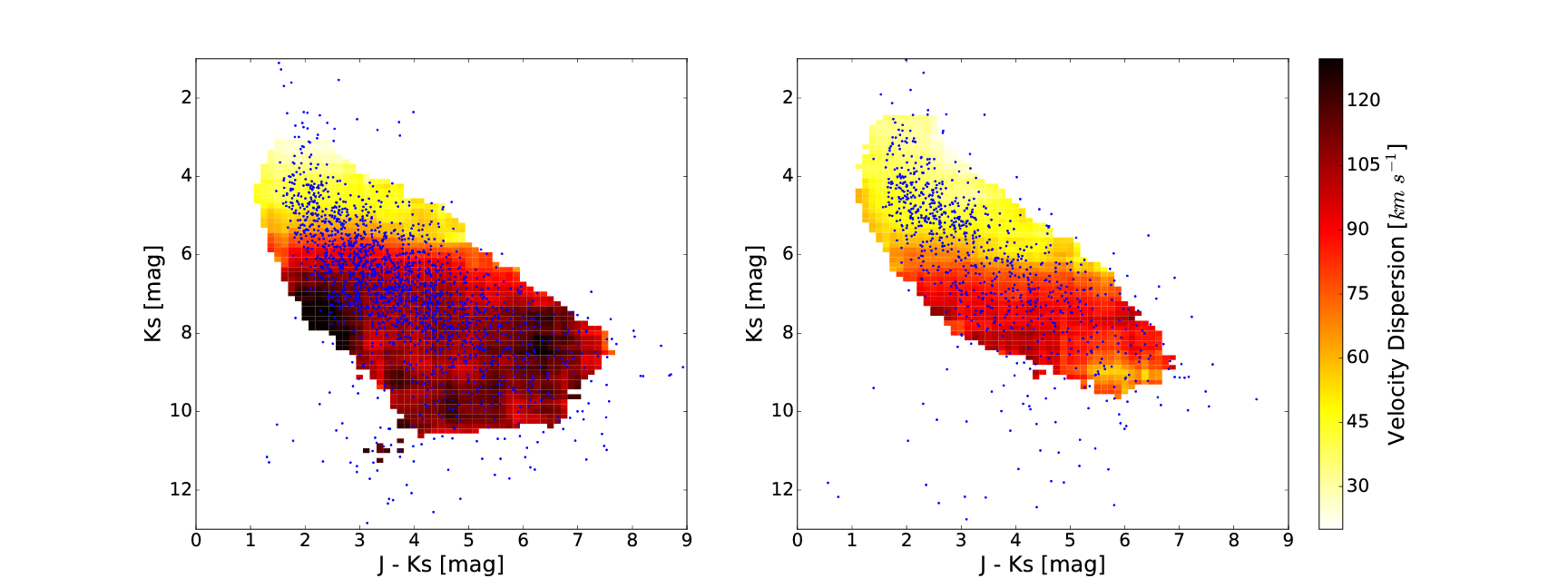}
   \caption{Total Galactocentric velocity dispersion as a color-map of BAaDE VLA (left) and ALMA (right) as a boxcar average in 2MASS $Ks$ vs $(J-Ks)$ space. Each colored box represents the velocity dispersion in a larger box around it of side length 1.25 magnitudes. The transition from low to high velocity dispersion around $Ks = 6$ is used to separate our sample into kinematically cold and hot populations, respectively.}
   \label{POP_SEP}%
\end{figure*}
\section{Data Acquisition} \label{sec:data}

We give a brief description of our sample selection here; the details are fully described in Sjouwerman et al. (2018) (in prep.).  The construction of the sample involves three broad steps: identification of likely maser sources from mid-IR photometry, follow-up with the VLA and ALMA to detect the masers with about a 50-75\% success rate, depending on longitude, observing conditions, etc., and the matching of the detected masers with 2MASS photometry to yield a color-magnitude diagram of maser sources with LoS velocities.

We initially select a candidate catalog based on mid-infrared photometry from MSX, as described in Sjouwerman et al. (2009).  The IRAS color-color diagram (Van der Veen and Habing 1988) enables a classification based on circumstellar shell thickness and opacity properties, for example, the targeting of sources with OH masers in high-mass-loss stars with thick opaque shells (e.g. te Lintel-Hekkert et al. 1991). However for sources in the Galactic plane, IRAS is confused due to its large (1 arcmin) beam; this led  Sjouwerman et al. (2009) to transform the IRAS color-color diagram into the mid-infrared MSX system.  Candidate infrared sources likely to be SiO masers can then be selected based on their loci in the MSX color-color plot, in regions where mass-loss and shell thickness are present, but much less extreme than for OH/IR stars. Our parent sample of $\sim$30,000 target infrared sources ($\sim$10,000 in the direction of the bulge) was selected from the MSX colors (MSX PSC version 2.3, Egan et al. 2003) as described by Sjouwerman et al. (2009). These targets are then observed with ALMA or the VLA for SiO masers at 43.1 GHz corresponding to $J = 1\rightarrow0 (v = 1)$, 42.8 GHz with $J = 1 \rightarrow 0 (v = 2)$, and 86 GHz with $J = 2 \rightarrow 1 (v = 1)$. The derived stellar velocities may vary with a few $\rm~km~s^{-1}$ from the exact stellar velocity as they are derived from the SiO maser peak emission velocity centroids (Sjouwerman et al. 2018).

The sources presented in this paper are a pilot of the full BAaDE survey ($\sim$2700 out of $\sim$20000). They are selected from the parent target list primarily by ease of observation. In the areas subtended by our sources in Fig.~\ref{POS}, an attempt was made to observe all available sources from the parent sample. However, bad weather conditions, lack of strong calibrators, or other observational problems could lead to a lack of uniform spatial coverage in the \textit{current} sample. For example, the lack of stars at $(l,b)\approx(-12,2)$ (see Fig.~\ref{POS}) has many targets in our parent sample, but have not been analyzed yet.

Nearly all of the stars in our sample are matched to 2MASS sources, giving access to $J$, $H$, and $Ks$ magnitudes. The sources are initially matched based on position comparison between the MSX and 2MASS catalogs, using a search radius of 5 arcsec.  For sources with more than one match within this radius, the brighter and redder sources were chosen. The typical positional offset between the 2MASS positions and an SiO maser position is $\sim 2$ arcsec.

Our sample lies brighter than and red-ward of the APOGEE red giant branch in the 2MASS $Ks$ vs $(J-Ks)$ color magnitude diagrams (see Fig.~\ref{CMD_1}).  Since these have been selected to have circumstellar envelopes based on their locus in the mid-IR MSX color-color plot, we presume that the extremely red $(J-Ks)$ colors are intrinsic and due to the circumstellar shells. For part of our analysis, we correct for interstellar dust using an extinction map from Nidever et al. (2012a).

In order to make our analysis comparable to earlier work, especially the {\sl BRAVA} survey, we correct our frame of reference to Galactocentric following Beaulieau et al. (2000). While the term ``Galactocentric" seems like it is referring to what an observer would see if placed at the center of the galaxy, this is not the case. These corrections place the observer at the location of Earth, but not moving relative to the center of the Galaxy. In this frame of reference, line-of-sight velocities and radial velocities are equivalent, as the origin of the coordinate system is at Earth's position. This velocity projection has been adopted in e.g. the BRAVA survey and gives a very clear visualization of the rotation of the Galactic bulge, as it tries to show something similar to what an observer outside our Galaxy would measure. Following Beaulieau et al. (2000) we adopt the circular velocity of the Local Standard of Rest as $220\rm~km~s^{-1}$ (Kerr \& Lynden-Bell 1986) and Earth's peculiar velocity as $16.5 \rm~km~s^{-1}$ toward $(l,b) = (53^\circ,25^\circ)$.


\begin{figure}
   \centering
   \includegraphics[width=\hsize]{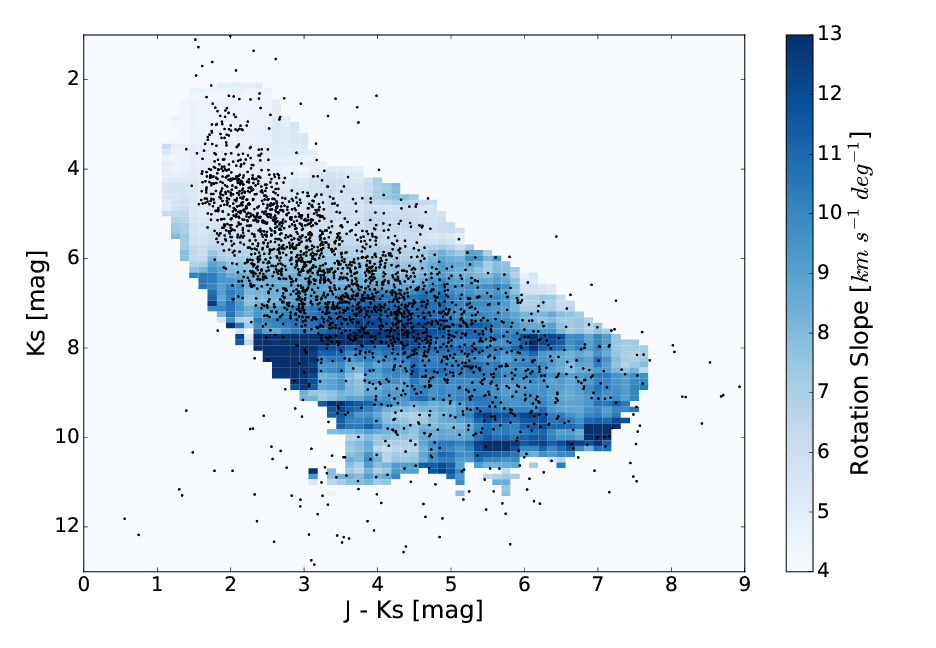}
   \caption{Rotation curve slope color-map of BAaDE VLA and ALMA stars as a boxcar average in 2MASS $Ks ~vs ~(J-Ks)$ space. Each box represents the slope in a larger box around it of side length 1.25 magnitudes. A transition from small to large slope values also happens around $Ks = 6$, supporting our separation into distinct kinematically cold (shallow slope) and hot (steeper slope) populations.}
   \label{POP_SEP_SLOPE}%
\end{figure}

\begin{figure*}[t!]
   \centering
   \includegraphics[width=\hsize]{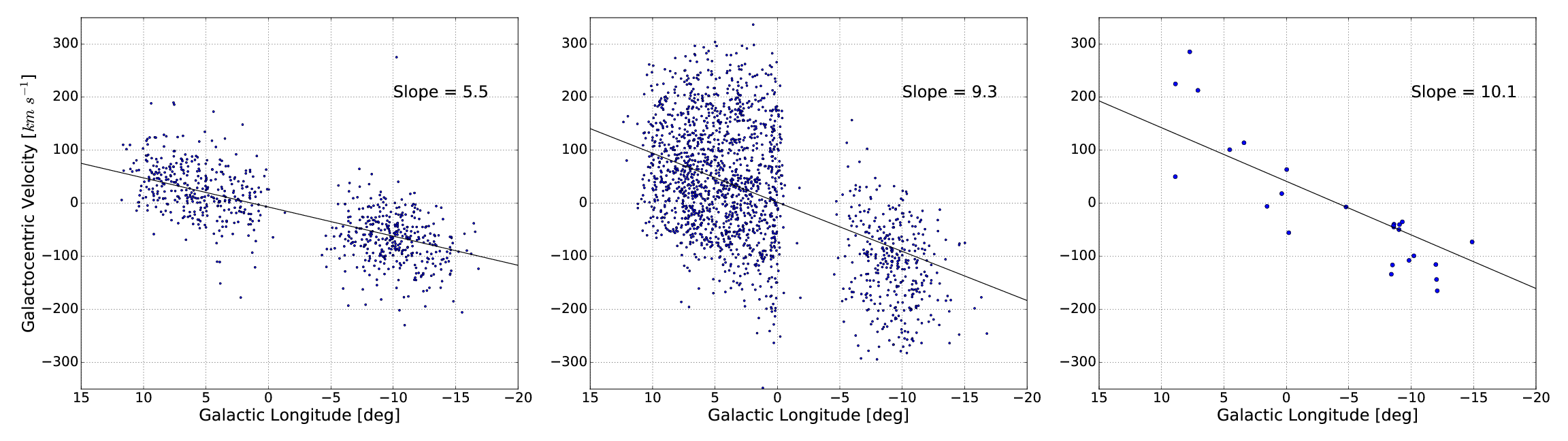}
   \caption{Radial velocity vs longitude for the kinematically cold (left), kinematically hot (center), and faint (right) BAaDE components. This figure highlights the kinematic differences between the three populations.}
   \label{SLOPE}%
\end{figure*}

\begin{figure*}
    \centering
    \includegraphics[width=\hsize]{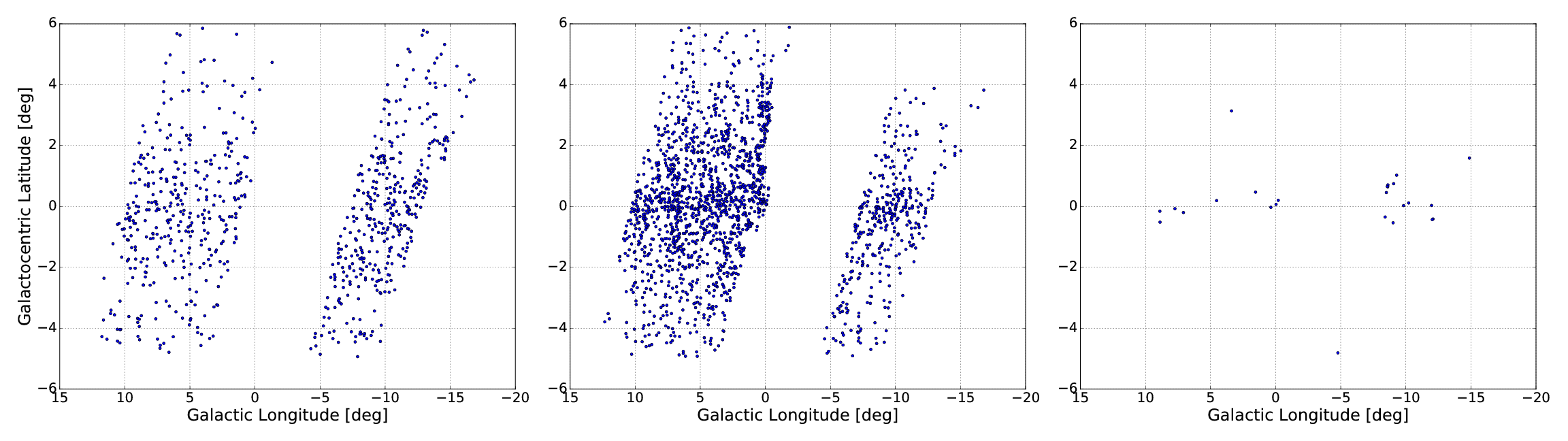}
    \caption{Spatial distribution of the three populations of stars. The cold kinematic population (\textit{left}) is more evenly spread in latitude, while the hot population (\textit{middle}) is more concentrated toward the Galactic plane. The faint population (\textit{right}) is only about $1^\circ$ wide (ignoring two outliers), which at a distance of 10 kpc (approximate distance to the opposite side of the bulge) corresponds to $\sim0.35$ kpc.}
   \label{POP_POS}%
\end{figure*}


\section{Kinematics Across the Color-magnitude diagram} \label{sec:pop_sep}

Fig.~\ref{POP_SEP} shows the velocity dispersion of BAaDE stars as a heat-map on the 2MASS color-magnitude diagram. A transition from low dispersion to high dispersion occurs at $Ks\approx 6$ for ALMA stars, and rather quickly at $Ks\approx5.5$ for VLA stars. These velocity dispersions were derived after subtracting the rotation curve of the sources (shown in Fig.~\ref{POP_SEP_SLOPE}), which shows a transition from a shallow to steeper rotation, also at $Ks\approx6$.
Thus, we label the fainter stars ($Ks > 5.5$ for VLA stars; $Ks > 6$ for ALMA stars) as the kinematically ``hot" population because of their higher velocity dispersion and faster rotation, and the brighter stars as the kinematically ``cold" population\footnote{ALMA hot population stars have a lower velocity dispersion than the VLA hot population stars by approximately $20\rm~km~s^{-1}$ on average; a possible explanation for this as well as the different transition magnitude is given in section \ref{sec:pop_analysis}.}. Fig.~\ref{SLOPE} (left and center panel) shows the separation of the two populations and compares their kinematics. Their positional differences are shown in Fig.~\ref{POP_POS} (left and center panel).

Selecting for the faintest stars in our sample ($Ks\gtrsim12$) gives a small velocity dispersion subsample similar to that of the cold kinematic population, but a slope more comparable to the hot population (see Fig.~\ref{SLOPE}). They are also more concentrated to the Galactic plane (b = 0) than the other two populations (see Fig.~\ref{POP_POS}). We will refer to this subset as the ``faint" population.

APOGEE stars are also plotted in Fig.~\ref{CMD_1} because of their high spatial overlap with most of our BAaDE stars. However, in color-magnitude space, BAaDE and APOGEE stars are nearly mutually exclusive; BAaDE stars are found notably brighter and redder compared with the bulk of the APOGEE sample.


\begin{figure*}
   \centering
   \includegraphics[width=\hsize]{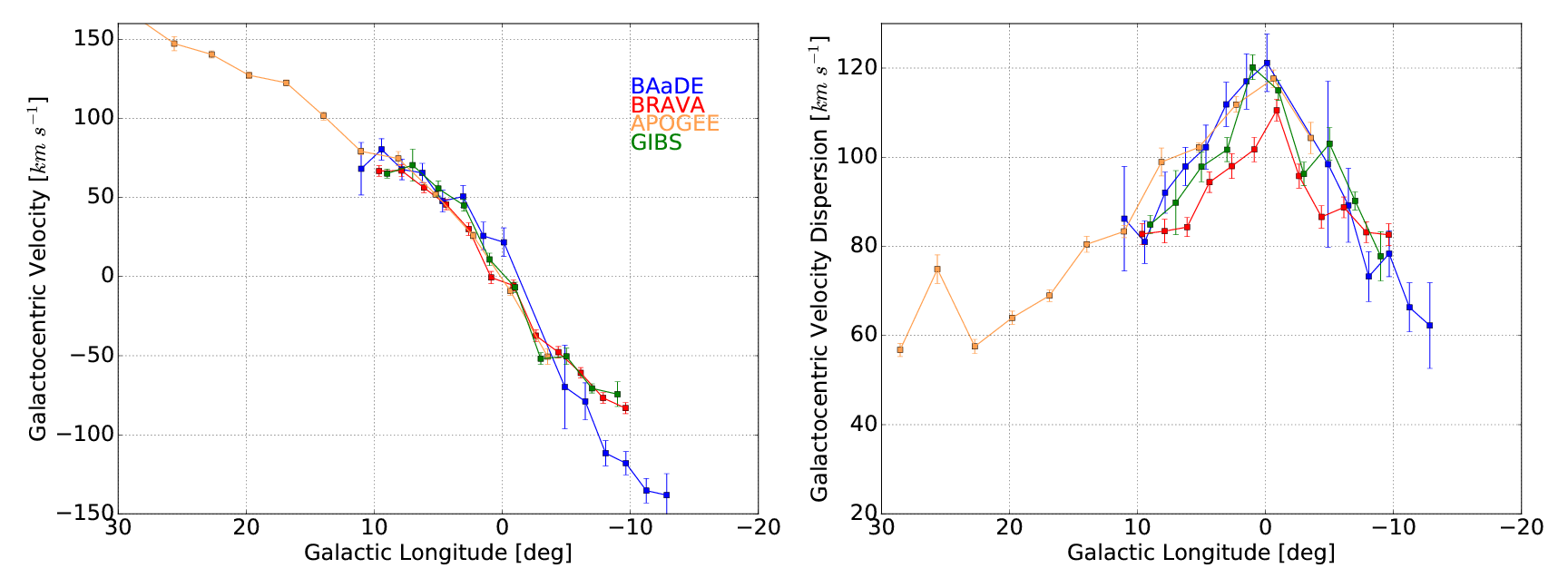}
   \caption{Binned rotation curves and velocity dispersion profiles for all four surveys. The hot kinematic populations of BAaDE and APOGEE (blue and orange) are shown here. APOGEE stars are chosen such that $T_{eff} \leq 4000$ to select for the APOGEE kinematically hot component. All surveys are consistent in rotation curves and velocity dispersions.}
   \label{HOT_COMP}%
\end{figure*}

\begin{figure}
    \centering
    \includegraphics[width=\hsize]{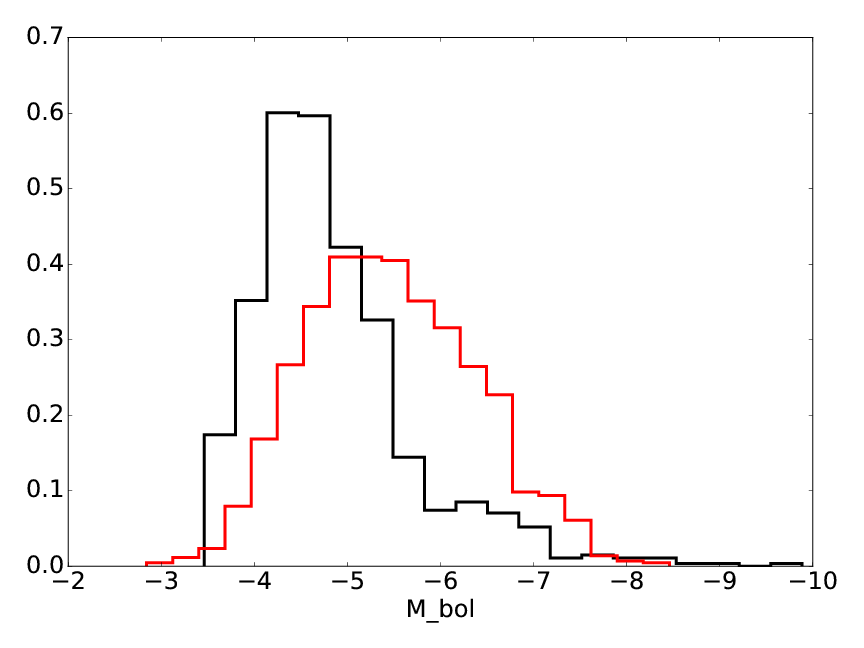}
    \caption{Red: Luminosity function of the hot kinematic population assuming: i) a common distance of 8.3 kpc (Gillessen et al. 2016), ii) Messineo (2004) SiO maser $Ks$-band bolometric corrections (pg. 60), and iii) dereddening using red clump stars from Nidever et al. (2012a). Black: Luminosity function of the cold kinematic population assuming: i) a common distance of 3.4 kpc (typical distance to a disk star towards the GC), and ii) Messineo (2004) SiO maser $Ks$-band bolometric corrections (pg. 60). Iben and Renzini (1983) determine a maximum bolometric magnitude of $M_{bol} \approx -7.2$; thus, sources in this figure with $M_{bol} < -7.2$ are likely closer than 8.3 or 3.4 kpc (for red and black, respectively), have an inaccurate bolometric correction, or have been insufficiently dereddened.}
   \label{LUM_FUN}%
\end{figure}

\begin{figure*}
   \centering
   \includegraphics[width=\hsize]{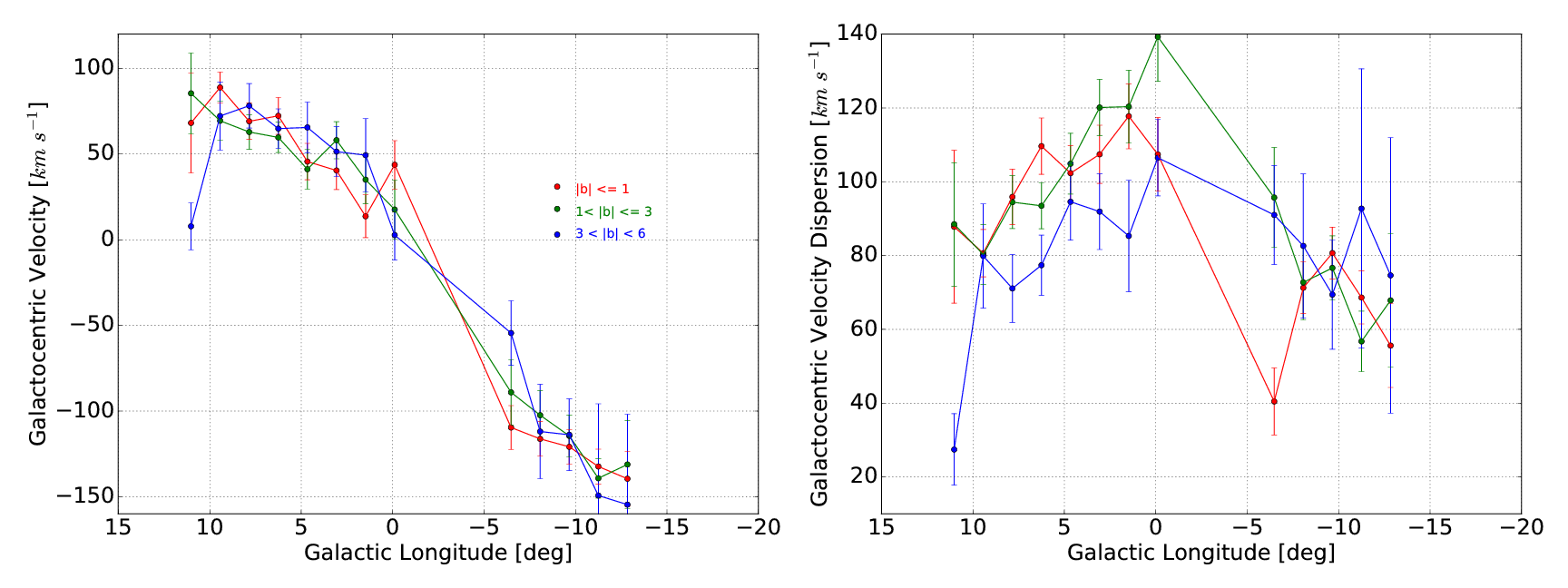}
   \caption{Rotation curves for stars for various latitudes. The left panel shows cylindrical rotation: no significant difference in rotation slope at different latitudes. This is a predicted property of a bar-like population (Shen et al. 2010). As reported in Rich et al. (2007), Kunder et al. (2012), and Zoccali et al. (2014), velocity dispersion decreases with latitude in the bulge. We cannot conclusively say the BAaDE dataset has this property (right panel). However, the final BAaDE dataset will have approximately 3 times the current number of sources and consequently smaller errors in the velocity dispersion. }
   \label{CYL}%
\end{figure*}

\begin{figure*}
   \centering
   \includegraphics[width=\hsize]{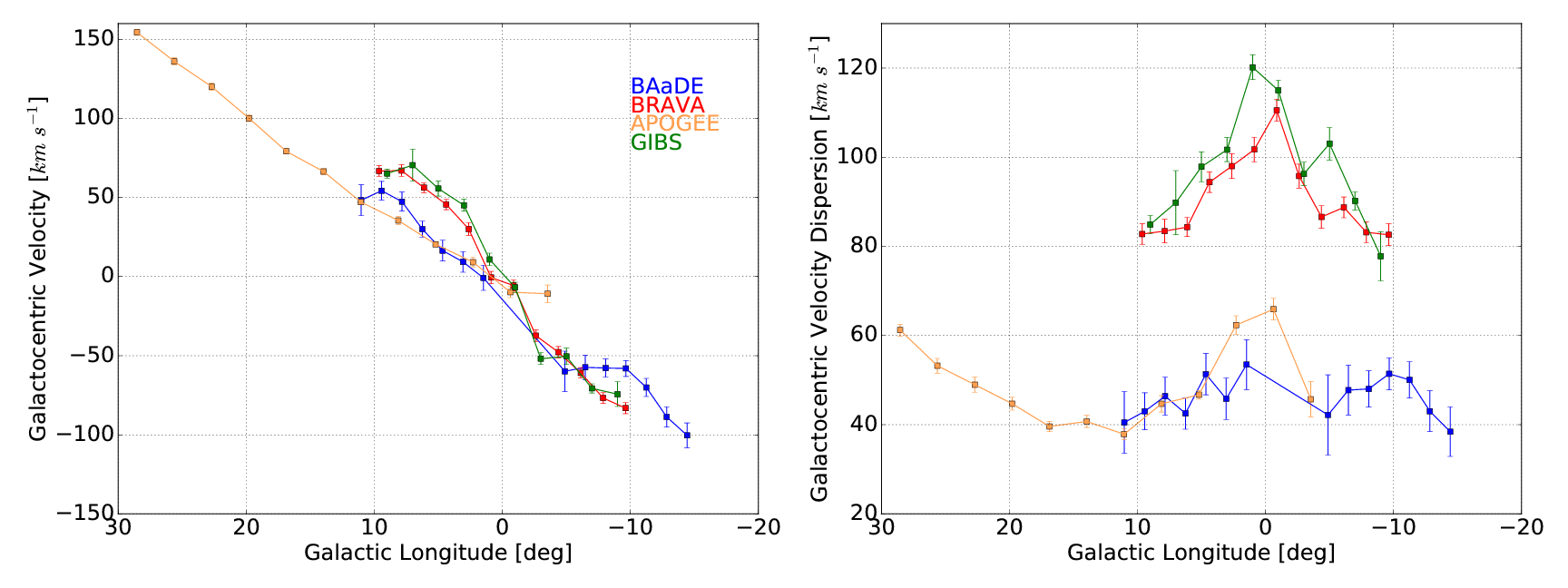}
   \caption{Binned rotation and dispersion curves for all four surveys. The cold kinematic populations for BAaDE and APOGEE (blue and orange) are shown here. APOGEE stars are selected such that $T_{eff} > 4000$ to select for the kinematically cold component. For APOGEE, we also exclude sources within 0.5 degrees of the galactic center because of the very large velocity dispersion there. Besides the obvious difference in dispersion relations between the kinematically cold and hot components, there is also a difference in slope on the left panel, indicating that the cold kinematic component consists of a more slowly rotating population, which is consistent with a foreground disk population.}
   \label{COLD_COMP}%
\end{figure*}

\begin{figure}
   \centering
   \includegraphics[width=\hsize]{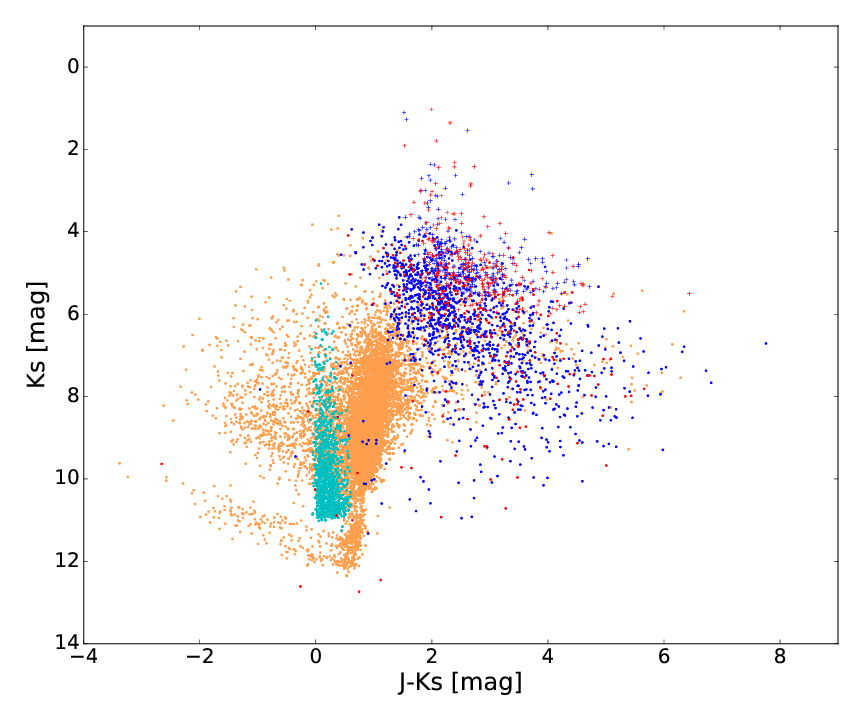}
   \caption{2MASS CMD dereddened using $Ks$-band red-clump extinction map (Nidever et al. 2012a). VLA: blue; ALMA: red; and APOGEE: orange (giants) and cyan (main sequence; not dereddened). Dots and crosses correspond to the kinematically hot (dereddened) and cold (not dereddened) populations, respectively.}
   \label{CMD_2}%
\end{figure}

\begin{figure*}
   \centering
   \includegraphics[width=\hsize]{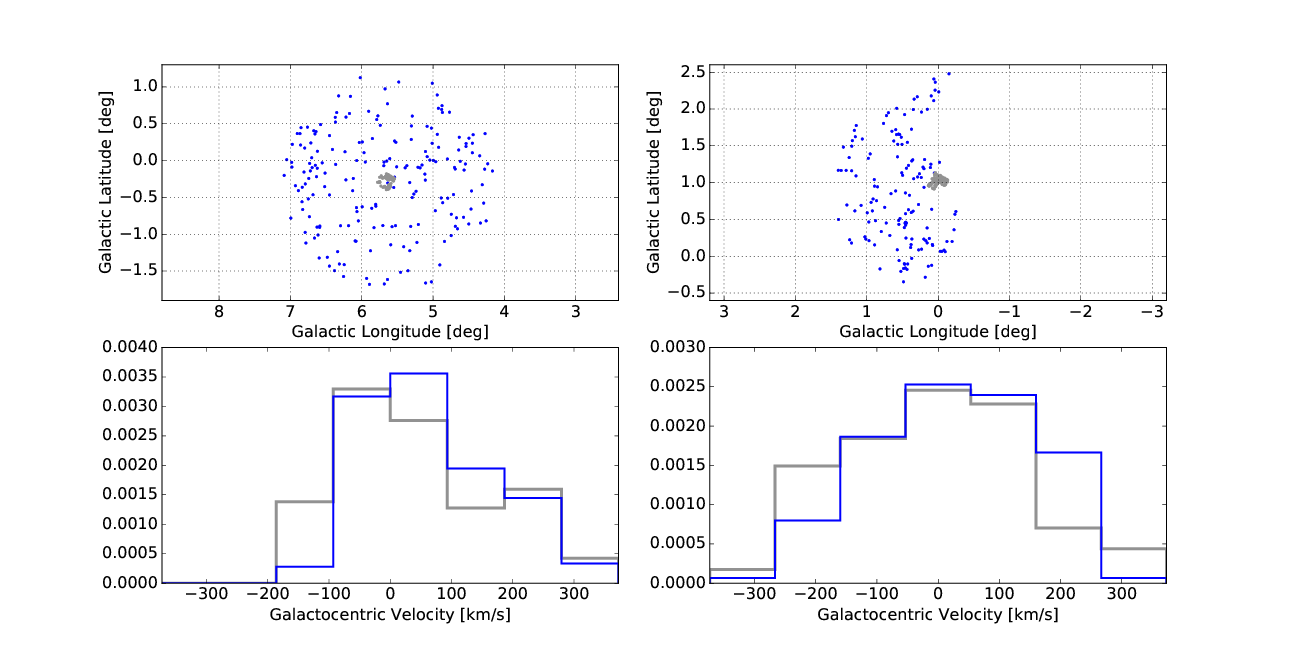}
   \caption{A line-of-sight velocity distribution comparison between two fields of Babusiaux et al. 2014 (gray) and BaADE sources (blue) within 1.5 deg of the Babusiaux sources. Top panels show position and lower panels show normalized velocity distribution histograms. The Kolmogorov\textendash Smirnov (K-S) test cannot conclusively say that the distributions are drawn from separate parent distributions, with a p-value of 0.156 and 0.089 (left and right panels).}
   \label{BAaDE_Babu}%
\end{figure*}

\begin{figure}
   \centering
   \includegraphics[width=\hsize]{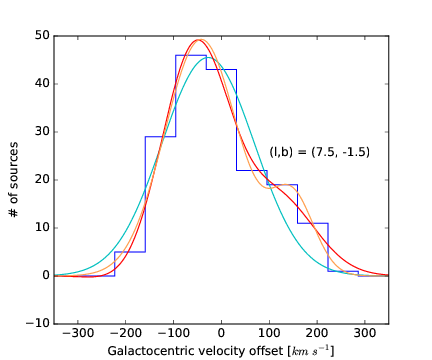}
   \caption{Histogram of the Galactocentric velocity offset of BAaDE VLA hot kinematic component stars in a 3x3 degree box centered at $(l,b) = (7.5,-1.5)$. ``Galactocentric velocity offset" refers to the difference in velocity from the linear rotation curve of slope $9.3\rm~km~s^{-1}~deg^{-1}$ shown in Fig.~\ref{SLOPE}. The cyan curve is a Gaussian; the orange curve is a double Gaussian; and the red curve is a Gauss-Hermite polynomial (van der Marel \& Franx 1993). There is an excess at $\sim 120 \rm~km~s^{-1}$ in in this frame (at $\sim 200\rm~km~s^{-1}$ in true Galactocentric), and the position on the sky is similar to that of the high velocity structure reported in Nidever et al. (2012b). We cannot, however, rule out that this is a manifestation of a skewed population as predicted in Zhou et al. (2017).}
   \label{200KMS_1}%
\end{figure}

\begin{figure*}
   \centering
   \includegraphics[width=\hsize]{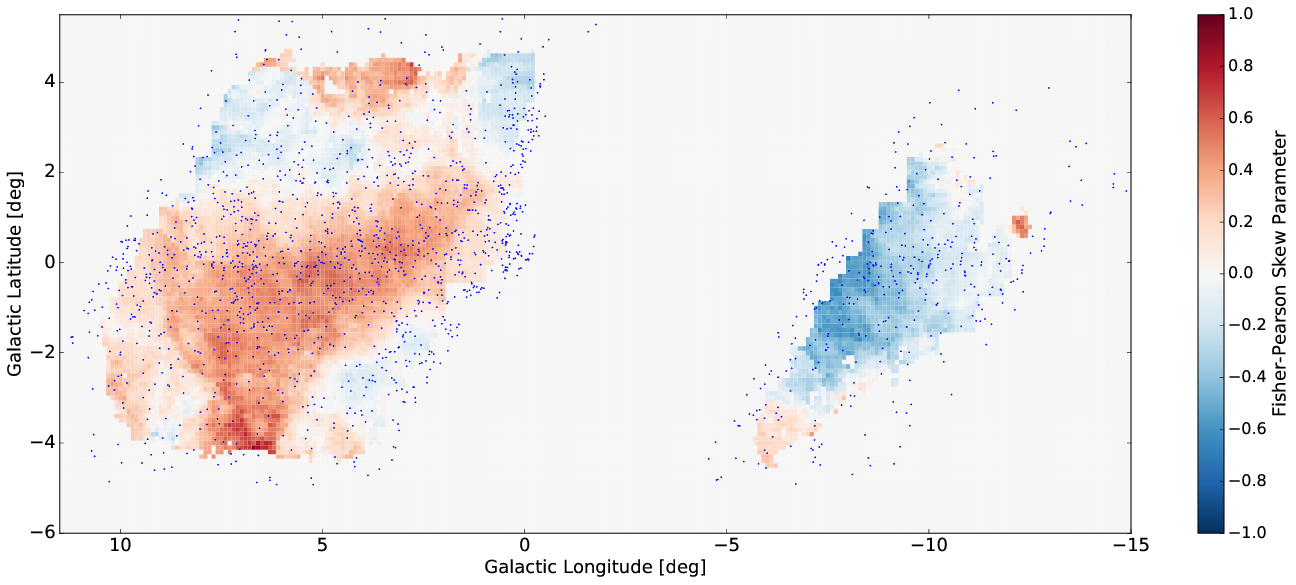}
   \caption{Skewness of the hot population as a function of position. This is a boxcar moving average plot, with the value of each point being the skew value of the stars in a $3^\circ$ diameter circle centered on that point. In general, the velocity distributions are skewed positive at positive longitudes, and negative at negative longitudes.}
   \label{SKEW_1}
\end{figure*}

\begin{figure*}
   \centering
   \includegraphics[width=\hsize]{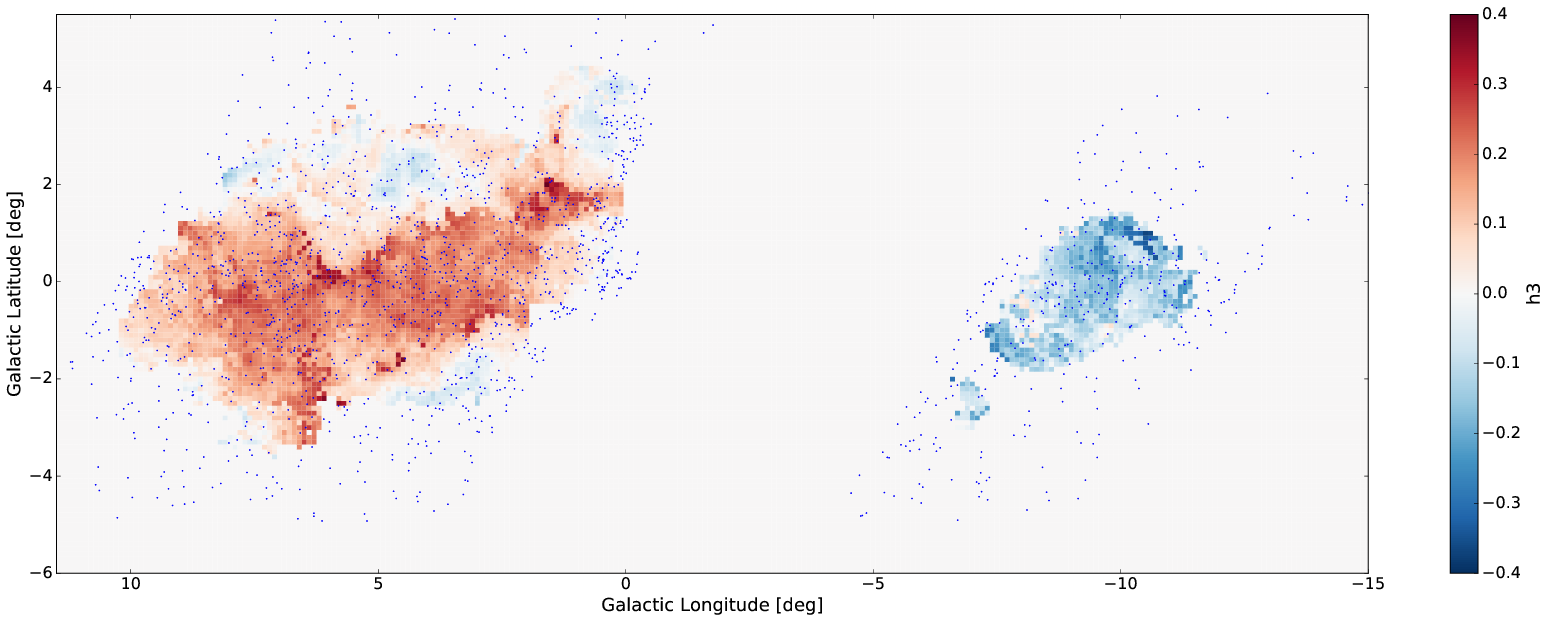}
   \caption{Gauss-Hermite ``h3" value (van der Marel \& Franx 1993) of the hot population as a function of position. This is a boxcar moving average plot, with the value of each point being the skew value of the stars in a $3^\circ$ diameter circle centered on that point. In general, the velocity distributions are skewed positive at positive longitudes, and negative at negative longitudes.}
   \label{SKEW_2}
\end{figure*}

\begin{figure}
   \centering
   \includegraphics[width=\hsize]{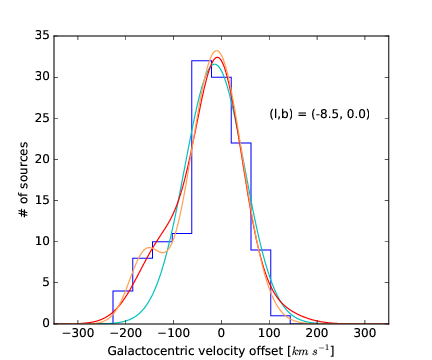}
   \caption{Histogram of the Galactocentric velocity offset of BAaDE ALMA hot kinematic component stars in a 3x3 deg box centered at $(l,b) = (-8.5,0.0)$. ``Galactocentric velocity offset" refers to the difference in velocity from the linear rotation curve of slope $9.3\rm~km~s^{-1}~deg^{-1}$ found in Fig.~\ref{SLOPE}. The cyan curve is a Gaussian; the orange curve is a double Gaussian; and the red curve is a Gauss-Hermite polynomial. There is a clear excess near $-150\rm~km~s^{-1}$ (near $-200\rm~km~s^{-1}$ in true Galactocentric).}
   \label{200KMS_2}%
\end{figure}

\section{Analysis of the Kinematic Populations} \label{sec:pop_analysis}
Given the separation of the masers into kinematic sub-populations, this section attempts to interpret the kinematic differences. We expect masers to exist in both the Galactic disk and bulge, so it is natural to assign the brighter sub-population to the disk, the fainter to the bulge, and the faintest to background disk stars. In this section, we provide additional evidence for these assignments, and compare our data to other bulge surveys.

\subsection{The Kinematically Hot Population}
Fig.~\ref{HOT_COMP} shows rotation curves for the hot component of BAaDE in comparison to the three other bulge surveys, plotting only the hot component for APOGEE\footnote{The hot component of APOGEE is defined in this paper as all stars with $T_{eff} < 4000K$ and no closer than 0.5 degrees to the galactic center (to get rid of extremely high velocity distribution stars). $T_{eff}$ was obtained from DR13, and it was calculated using radial velocity templates.}. We assign the BAaDE hot component to the bulge primarily because of the agreement in both rotation and velocity dispersion with BRAVA, GIBS, and the APOGEE hot component. As reported in Rich et al. (2007), Kunder et al. (2012), and Zoccali et al. (2014), velocity dispersion decreases with latitude in the bulge, explaining the slightly lower velocity dispersion of BRAVA compared to the other surveys; the BRAVA survey is at higher average latitude.

Fig.~\ref{LUM_FUN} shows a luminosity function of the hot kinematic population assuming: i) a common distance of 8.3 kpc (Gillessen et al. 2016), ii) Messineo (2004) SiO maser $Ks$-band bolometric corrections using 2MASS-MSX colors (pg. 60), and iii) dereddening using the Nidever et al. (2012a) red clump extinction map. The peak of bolometric luminosity is $M_{bol} \approx -5.5$, which reaches that of the brightest M giants in the bulge (Frogel and Whitford 1987).  The full extent of the histogram reaches $M_{bol}<-7$, and the whole population would appear to be significantly more luminous, and thus younger, than the intermediate age Magellanic Cloud clusters and Bar West in the LMC (Frogel, Mould, \& Blanco 1990).  However, this histogram is an approximation. Because the actual distance to bulge stars ranges from approximately 5 kpc to 12 kpc, the assumption of a common distance of 8.3 kpc broadens the histogram by $\sim 1$ mag in each direction. Iben and Renzini (1983) determine a maximum bolometric magnitude of $M_{bol} \approx -7.2$; thus, sources in this figure with $M_{bol} < -7.2$ are likely to be closer than 8.3 kpc. For M giants with such extreme mass loss, it is important to undertake more careful modeling of the spectral energy distributions to model the bolometric magnitudes.  Although the peak of the luminosity function is extreme, old metal rich populations with the age of globular clusters can have AGB stars reaching $M_{bol}=-5$ and brighter (Guarnieri et al. 1997).  Further, some fraction of these luminous AGB stars might be the progeny of blue straggler mergers. We also plan to do a full kinematic/spatial distribution model to derive the final luminosity function.  Nonetheless, taken with the evidence of long period Miras in the bar, the luminosity functions would appear to support the SiO masers being the progeny of an intermediate age population. 

ALMA stars in Fig.~\ref{POP_SEP} (right panel) appear slightly different than the VLA stars (left panel). They transition from a cold to hot population at $Ks\sim6$ rather than the VLA sample's $Ks\sim5.5$. The transition magnitude difference could be a manifestation of the fact that the bar is more distant on that side of the Galactic Center, causing the transition from disk population to bulge population to happen at a higher magnitude (farther distance). Also, ALMA kinematically hot stars ($Ks>6$) have a lower velocity dispersion than the VLA's by approximately $20\rm~km~s^{-1}$ on average. For the bulge, velocity dispersion decreases with longitude (e.g., Rich et al. 2007, Zoccali et al. 2014), and indeed the ALMA stars are (on average) farther from the Galactic Center along the plane than the VLA stars (see Fig.~\ref{POS}). The velocity dispersion of the ALMA stars is consistent with the other bulge surveys when taking into account longitude (see Fig.~\ref{COLD_COMP}, right panel).

Many bar models of the bulge predict cylindrical rotation (Shen et al. 2010), i.e. the rotation curve is independent of height above the plane (latitude). Fig.~\ref{CYL} shows that the BAaDE hot kinematic population does exhibit cylindrical rotation, consistent with it being bar-shaped, rather than an elliptical bulge. On these grounds, we assign the hot kinematic population to be majority bulge stars.

\subsection{The Kinematically Cold Population}

Fig.~\ref{COLD_COMP} shows the cold components of BAaDE and APOGEE plotted with the other bulge surveys. The shallower rotation curves and dramatically lower velocity dispersion for the cold components imply a different population, likely foreground disk stars.

We assign the cold population to the foreground disk based on four main pieces of evidence:
\begin{enumerate}
    \item The velocity dispersion of the cold population ($\sim 50\rm~km~s^{-1}$) is most consistent with a disk population of up to ($\sim7 \rm~Gyr$) and inconsistent with any purely young population (Nordstrom et al. 2004).  
    \item Historically, the brightest bulge giants have $M_{bol} =-5.5$ corresponding to $M_K \sim -9.5$ (Frogel \& Whitford 1987).  Placing the brightest bulge giant at the nearest end of the bar ($\sim 5$ kpc away), it would have an apparent $K_0 = 4$. All stars brighter than this are likely to be disk stars closer to the Sun.  Allowing for $Ks$-band extinction values of 1-2, this is close to our our observed transition of $Ks\sim6$. 
    \item Fig.~\ref{LUM_FUN} shows a luminosity function of the cold kinematic population assuming: i) a common distance of 3.4 kpc (typical distance to a disk star towards the GC), and ii) Messineo (2004) SiO maser $Ks$-band bolometric corrections using 2MASS-MSX colors (pg. 60). The peak of bolometric luminosity is $M_{bol} \approx -4.5$, consistent with their being luminous and possibly intermediate age AGB stars (Habing 1996) in the disk. If the population were at 8 kpc, most of this population would exceed maximum AGB luminosity of $M_{bol} \approx -7.2$ (Iben and Renzini 1983).
    \item We create simple model of a foreground disk of stars with a flat rotation curve, circular orbits, a velocity dispersion of $50\rm~km~s^{-1}$, an exponential disk with scale length 3 kpc, and an inner radius of 4 kpc. This model predicts a rotation slope of $\sim7\rm~km~s^{-1}~deg^{-1}$. This is consistent (within a slope value of $1.5\rm~km~s^{-1}~deg^{-1}$) with the cold kinematic component (slope = $5.5\rm~km~s^{-1}~deg^{-1}$; see Fig.~\ref{SLOPE}), and substantially smaller than the rotation slope of the hot kinematic component assigned to the bulge. More elaborate models are planned for future work as ours is a 2-D model and does not take into account the sensitivity limits of MSX, but rather assumes all stars simulated are observed. 
\end{enumerate}

For both the hot and cold components, the rotation curve slopes and velocity dispersion of BAaDE and APOGEE are very similar to one another (see Fig.~\ref{HOT_COMP} and \ref{COLD_COMP}), suggesting that they are probing the same population of stars despite being in different areas of color-magnitude space (see Fig.~\ref{CMD_1}). 

The spatial distribution (Fig.~\ref{POP_POS}) of the cold population differs from that of the hot component. In both cases, there is a concentration of stars toward the plane, but the density of the cold population is not as peaked at the plane. One might expect such a distribution if the disk population is nearby compared to the bulge.

On these grounds, we assign the majority of the cold population to be foreground thick-disk stars. 

When the dereddening is applied (Nidever et al. 2012a) to the kinematically cold population, it is immediately clear that they are closer to the Sun than the reddening sheet because of their unphysically blue colors. They are too deredenned by the extinction map.  Thus, the kinematically cold population is not deredennend in this paper. 

Fig.~\ref{CMD_2} shows a CMD that has been dereddened for the kinematically hot population only using the extinction map by Nidever et al. (2012a). Our sources now occupy a more compact region of color-magnitude space, but are still separate from the APOGEE stars.

While the hot kinematic component is likely very old evolved stars, it is likely that the cold kinematic disk component discussed in this section is associated with more massive evolved stars, because of the typically younger age of disk stars.  However, while our separation is robust, it is not perfect, and some cross-contamination of disk and bulge may be present and affect the LOSVD (Section \ref{sec:skew}).

The kinematically cold population in Fig.~\ref{POP_POS} (left panel), appears
to show an irregular distribution of stars, with a possible decline in numbers
for $ \vert b\vert \approx 3^\circ$. There is no immediately obvious explanation for this; if this is an effect of our selection from MSX, a similar deficit would be expected but is not observed in the kinematically hot population.  As the gap in data at $-5^\circ<l<0^\circ$ will be filled, it will be interesting to see if this persists.

\subsection{Faint Population}
Fig.~\ref{SLOPE} shows that the faint population's rotation slope is similar to that of the hot population, but its velocity dispersion is even less than that of the cold population.

The low-latitude spatial distribution of the faint stars in Fig.~\ref{POP_POS} suggests they are on average farther away than the cold or hot population, or they are more confined to the plane. The angular thickness of the faint population is $\sim$2 deg. At a distance of $\sim$10 kpc (an approximate background disk distance), that corresponds to a vertical extent of $\approx 350$ pc, which is consistent with the thin disk vertical scale length (Bland-hawthorn \& Gerhard 2016).

We create a simple model of a background disk of stars with inner and outer radii of 3 and 5 kpc from the GC, a flat rotation curve, circular orbits, an exponential disk with a scale length of 3 kpc, and a velocity dispersion of $50\rm~km~s^{-1}$. This model predicts a rotation slope of $\sim 8 \rm~km~s^{-1}~deg^{-1}$. This is consistent (within a slope value of $2\rm~km~s^{-1}~deg^{-1}$) with the faint population (see Fig.~\ref{SLOPE}). More elaborate models are planned for future work as ours is a 2-D model and does not take into account the sensitivity limits of MSX, but rather assumes all stars simulated are observed.

We believe that the magnitudes and rotation curve support our proposal that the ``faint'' population corresponds to members of the thin disk on the distant side of the bar; additional data will help to constrain better the nature of these stars as the current numbers are small.

Fig.~\ref{POP_POS} shows the spatial distribution of the three populations. Qualitatively, we can understand the spatial distribution of the cold kinematic population as corresponding to a relatively nearby population, while the bulge population at $\sim 8$ kpc appears more concentrated to the plane both because it is more distant and because of the high density of stars in the bulge.  The apparent concentration to the plane of the distant disk stars in the right-hand panel reflects their relatively small scale height and their greater distance.

\subsection{Discussion}
Ours is not the first study to identify two kinematic populations of maser sources toward the bulge.  Lindqvist, Habing, \& Winnberg (1992) found two OH/IR maser populations in the central 100 pc of the Galaxy, when the masers were segregated by shell outflow velocity.  At first glance, our SiO masers  exhibit similar characteristics, with an older population of greater velocity dispersion, and a candidate younger, rotating population of lower velocity dispersion.  However, our sample has SiO masers and not OH/IR stars; the latter are a shorter lived evolutionary phase of high luminosity.  In addition, the OH/IR population of the Galactic Center has been identified as distinct on two counts.  First, as discussed in Rich (2013), Sjouwermann et al. (1999) found the distribution of OH maser expansion velocities to peak at $20 \rm~km~s^{-1}$ in the Galactic Center, in contrast to the lower-value distribution seen the bulge. Second, Blommaert \& Groenewegen (2007) found the longest-period Miras concentrated to the central 50 pc.  All agree that these candidates for a younger population are present {\it in situ} in the Galactic Center.  Our candidate cold kinematic population of masers are so bright that only by adopting the mean 3.4 kpc distance can their luminosities be lower than the maximum allowed for AGB stars (Iben \& Renzini 1983) (see section \ref{sec:pop_analysis}).  We also note that the intrinsic colors of members of the kinematic hot (bulge-like) population extend to redder $(J-Ks)$ colors.  We conclude that our maser populations, while initially selected by kinematics, appear also to exhibit different physical properties and occupy different populations and mean distances (disk vs bulge); they are quite distinct from the two populations of OH/IR stars that are physically near the Galactic Center identified by Lindqvist, Winnberg, and Habing (1992).  

The Galactic Center has long been known to have a unique stellar population, hosting both a young stellar population and an extended star formation history  (e.g., Morris \& Serabyn 1996; Figer et al. 2004).  There is also an underlying population of stars with age $>$ 10 Gyr, supported by RR Lyrae stars in the central cluster (Dong et al. 2017) and red clump stars that are likely to be very old (Figer et al. 2004).  Based on this extended and complex star formation history, the presence of both an extraordinary population of OH/IR stars and long-period Mira variables would then be expected.  

Although there is some evidence for an intermediate-age stellar population in the Galactic bulge/bar based on microlensed dwarfs (Bensby et al. 2017) and long-period Miras (Catchpole et al. 2016), our study poses a significant challenge. The large velocity dispersion of our bulge SiO maser population is difficult to reconcile with a large fraction being intermediate age, formed in a disk \textit{already} in the presence of a bar-like bulge and heated to the present dispersion over $\sim 7$ Gyr timescales (Nordstrom 2004). Even a scenario involving formation in the plane and vertical diffusion due to encounters with molecular clouds, such as was invoked to account for the dynamics of OH/IR stars in the central few degrees (Kim \& Morris 2001), cannot produce the requisite high velocity dispersion of the bulge SiO maser stars.  

In fact, the high velocity dispersion of our bulge SiO maser population is similar to that of the low latitude red clump bulge stars of Babusiaux et al. (2014); Fig.~\ref{BAaDE_Babu}.  The red clump is a population of helium burning stars on the horizontal branch with age $>1$ Gyr, but such stars can have age ranging from 1-10 Gyr.  We see no hint of kinematic substructure in either the SiO masers or the clump.  Both populations peak at $\sigma = 140\rm~km~s^{-1}~deg^{-1}$ near $l=0$, far higher than any young stellar population.  If the SiO maser AGB stars are intermediate age, it would be reasonable to speculate they are $\sim 8-10 $ Gyr old or on the old side of intermediate age.
Such an age might allow for luminous AGB stars at high metallicity, and the the oldest bulge populations (RR Lyrae and short period Miras) might still be 1-2 Gyr older and not show the spatial distribution of the bar.


\section{Line of Sight Velocity Distribution} \label{sec:skew}

N-body simulations (Bureau and Athanassoula 2005; Iannuzzi and Athanassoula 2015) show higher order moments of the line of sight velocity distribution (LOSVD) can provide an interesting constraint on bar models. The kinematically hot component of the BAaDE sample is of special value in this regard, as it is unlikely to be contaminated by foreground disk stars; we also have a significant sample of stars at $l<0$, enabling us to probe regions of the bulge where the LOSVD is predicted to skew toward negative velocity.

The hot kinematic population exhibits a significant skew (see Fig.~\ref{200KMS_1} for an example of a population exhibiting a significant skew). The value of this skewness\footnote{Skewness is calculated as the Fisher-Pearson coefficient of skewness}, and the Gauss-Hermite ``h3" parameter\footnote{A parameter describing skew (van der Marel \& Franx 1993).}, are plotted as a function of position in Fig.~\ref{SKEW_1} and Fig.~\ref{SKEW_2} respectively. There is strong positional agreement between the two methods. Similar skewness is also seen in APOGEE at positive latitudes in Zasowski et al. (2016). In general, our velocity distributions are skewed positive at positive longitudes, and negative at negative longitudes, consistent with triaxial bar (see Bureau and Athanassoula 2005; Iannuzzi and Athanassoula 2015). The cold kinematic population does not exhibit a systematic skew like the hot population. However, when plotting the hot and cold kinematic components together, there is a \textit{larger} skew than the hot kinematic population alone. This is because the cold population (while not skewed its-self) has a lower rotation slope (see Fig.~\ref{SLOPE}). Adding the cold population to the hot population will move the mean of the combined velocity distribution towards lower velocities at larger longitudes, creating a positive skew. At negative longitudes, this would have the opposite effect: creating a negative skew. This is in fact the same skew direction that the hot kinematic population shows on its own. Thus, the hot component's skew could be a manifestation of contamination by cold kinematic stars that are faint enough for the $Ks$ magnitude cut to place them in the hot component, an effect that certainly is happening on some scale due to our rather unsophisticated method of separating the two populations. However, it is unlikely such a contamination could cause a skew as great as that seen in Fig.~\ref{SKEW_1} given how well the hot kinematic population matches the rotation slope and velocity dispersion of the bulge. If the hot kinematic population truly is a majority bulge population as we predict, it is intrinsically skewed. Additional data and a more refined population separation technique may resolve this degeneracy.


\section{Evidence of Kinematic Substructure} 
\label{sec:200KMS} 

Nidever et al. (2012b) reported a high velocity group of stars in the Milky Way bulge. These stars comprise approximately 10\% of the stars in their sample, and have radial velocities of approximately $200 \rm~km~s^{-1}~deg^{-1}$. This high velocity group extends from $l = 4^\circ$ (the innermost part of their sample) to $l = 14^\circ$, close to the plane ($|b| < 5^\circ$). We find a similar high velocity group ranging in prominence between longitudes if 6 to 9 degrees (the outermost part of our current preliminary sample), also close to the Galactic plane ($|b| < 5^\circ$), and with a characteristic velocity of $200\rm~km~s^{-1}~deg^{-1}$ (see Fig.~\ref{200KMS_1}). However, none of our VLA fields show a \textit{distinct}, well-separated secondary Gaussian feature; our data are also consistent with our stars exhibiting a skewed velocity distribution, like those presented in Zhou et al. (2017). As discussed in section \ref{sec:skew}, these skews could \textit{also} be caused by a contamination by disk sources, though this is unlikely.

On the other side of the Galactic Center, sources observed with ALMA probe a previously unexplored area in comparison to the other large bulge surveys (see Fig.~\ref{POS}). These data \textit{do} show a rather distinct secondary Gaussian feature at high velocity, most prominent near $(l,b) = (-8.5,0)$ (see Fig.~\ref{200KMS_2}) with a characteristic velocity of $-200 \rm~km~s^{-1}$. We propose these peaks as a candidate symmetric counterpart to the high velocity group found in Nidever et al. (2012b), but again, this double Gaussian could be the result of a skewed population suffering from low number statistics.

The high-velocity peak appears to be present only in the hot population, suggesting that this feature belongs to the bulge rather than the disk.

Debattista et al. (2018) predict a number of observational properties of a nuclear disk that are consistent with our LOSVDs. They predict the high velocity peak to be more prominent on the $l<0$ side, and the peak to be located at larger $|l|$ on the $l<0$ side. We see both of these properties in figures \ref{200KMS_1} and \ref{200KMS_2}. Debattista et al. (2018) also predict that the peak on the $l<0$ side should be at lower $|v|$ than on the $l>0$ side. However, the large uncertainty in the velocity position of the peaks prevent us from testing this prediction with the current sample. We will continue to collaborate with theorists as our full data set will provide good constraints on bulge/bar models.


\section*{Conclusions} \label{sec:Conclusion}
Our survey of SiO maser sources toward the Galactic bulge/bar allows us to probe the kinematics of the bulge/bar into the Galactic plane regardless of extinction, has isolated a sample of mass losing AGB stars, and extends a kinematic survey to the Galactic plane at negative Galactic longitude.  We present results from $\sim$2700 radio sources toward the Galactic bulge/bar, $\sim 20\%$ of the anticipated final sample of detected sources in the BAaDE survey; a detailed description of methods and a catalog are given in Sjouwerman et al. (2018) (in prep.). 

Using the MSX positions, we match our radio sources to 2MASS photometry, revealing clear evidence for luminous, mass losing stars.  The selection based on mid-infrared color and detected maser emission gives a sample with characteristics quite distinct from conventional optical/near-IR samples, and is excellent at isolating a population of mass-losing AGB stars.  The color-magnitude diagram of radio-matched stars is a broad swath brighter than the RGB tip and extending to $(J-Ks)>8$; there are almost no matches with the APOGEE sample.  Noting a break in the kinematics for $Ks > 5.5$, we identify two distinct kinematic populations: a population of higher velocity dispersion $\sigma =\rm 100 \rm~km~s^{-1}$) that we associate with the bulge/bar, and a lower velocity dispersion ($\sigma= 50\rm~km~s^{-1}$) foreground population that we assign to the disk.  We also argue that the faintest stars in our sample ($Ks>12$) correspond to disk stars seen on the far side of the bar.

Making the analogy with other optical/infrared selected bulge surveys, we conclude that the kinematically hot population of maser sources is associated with the Galactic bulge/bar.  We show that the rotation curve and velocity dispersion profiles of this population match those of e.g. the BRAVA, APOGEE, and GIBS populations.  We emphasize that although prior studies such as Lindqvist, Winnberg, and Habing (1992) have found kinematic subsets in maser populations, these studies were confined to the inner 50 pc of the galaxy and used OH/IR stars.  In contrast, we argue for a widespread dichotomy between bulge-like and thick-disk-like populations in the SiO maser sample; this is a new result.

For an approximate distance of 8.3 kpc, we estimate their bolometric magnitudes to be near $\sim-6$, making the SiO maser population among the most luminous evolved stars seen in the bulge.  The luminosities of these stars and their striking red colors associate these with mass losing AGB stars and Mira variables.  The high luminosities and red colors of these stars appear most consistent with an age $<10$ Gyr.  We note however that the velocity dispersion of our hot kinematic component is identical to the $\sigma = 145 \rm \rm~km~s^{-1}$ for the red clump stars near $l=0,b=1^\circ$ reported by Babusiaux et al. (2014); neither red clump stars nor SiO masers show any hints of kinematic substructure.

There is at present considerable controversy concerning the age distribution of stars in the bulge, with Bensby et al. (2013, 2017) arguing from analysis of micro-lensed dwarfs that the bulge must contain a substantial population of metal rich stars younger than 10 Gyr.  This appears to be reinforced by the most luminous stars, like those that we consider in this study.  Van loon et al. (2003) also argue for a widespread population of bulge stars younger than 7 Gyr, with an extended star formation history.  Catchpole et al. (2016) argue that the Miras of period $>400$ days are in the bar while shorter period Miras are not.  It is noteworthy that Blommaert \& Groenewegen (2007) find the longest period Miras to be confined to the inner $0.1^\circ$ or 15 pc. 
While we cannot consider all of the lines of evidence concerning the age of the bulge in this paper, there is clearly tension between the $\sigma \sim 140\rm \rm~km~s^{-1}$ found for our SiO maser sources (and for the red clump stars) and any plausible disk population of age $\sim 1$ Gyr.  New wide field imaging of the bulge with the CTIO DECam facility (The Blanco DECAm Bulge Survey; Rich et al. 2018 in prep) will place limits on main sequence stars in the bulge $<1 \rm~Gyr$ old.  One way out of this conundrum is the possibility that the suprasolar metallicity allows for stars with $\sim 10$ Gyr ages and high turnoff masses to reach higher than expected AGB luminosities.  The bulk of the mass in the bar may have formed in a burst $\sim 7-10$ Gyr ago corresponding to a formation redshift of $z\approx 1$ for the bar; the fraction of bars is seen to decline with redshift, but 20\% of luminous galaxies at z=0.84 are seen to host bars (Sheth et al. 2008). 

Our maser sample appears to follow the general trend of rising velocity dispersion of the metal rich component toward the Galactic plane (Babusiaux et al. 2016) that would appear to be inconsistent with the SiO maser population being the progeny of a much younger stellar population.   We also emphasize the similarity of our high velocity dispersion kinematic subsample to all of the major bulge surveys, especially APOGEE, while also recalling that the evidence from analysis of deep, proper motion-cleaned HST photometry does not strongly support the presence of a widespread intermediate age bulge population (e.g. Kuijken \& Rich 2002, Zoccali et al. 2003, Clarkson et al. 2008, Gennaro et al. 2015). The apparently extreme luminosities of the SiO masers that seem in tension with the implications from the kinematics pose an issue that should be addressed and resolved.  Because our maser sample has excellent radial velocity measurements and a clearly identified evolutionary state, the further study of this sample will offer some powerful constraints on the age distribution and formation of the bulge.

Examining the line of sight velocity distribution (LOSVD), we detect skew moments in the hot kinematic population.  We also find a high velocity ($200 \rm~km~s^{-1}$) substructure may correspond to that reported by Nidever et al. (2012b), and a new candidate high velocity substructure ($-200 \rm~km~s^{-1}$) at $l<0^\circ$.  For the first time, we see the LOSVD display skew toward negative radial velocities at negative galactic longitude, and skew toward positive radial velocity at positive galactic longitude.

VLBI observations of a subset of our sample are being planned in order to obtain full 3D velocities and parallax distances; for a substantial number of our disk sources, distances and proper motions will eventually become available via the Gaia experiment.  Our team also is modeling the spectral energy distributions of the SiO maser sources, which should provide improved estimates of bolometric luminosity.  The comparison of these with stellar evolution models offer the real hope that the ages of the maser population can be better understood.



\section*{Acknowledgements} \label{sec:Ack}
We would like to thank the referee of this paper for insightful comments, and constructive guidance in making this a better article.

The BAaDE project is funded by National Science Foundation Grant 1517970/1518271.
This paper uses data products obtained with instruments run by the National Radio Astronomy Observatory (NRAO): the VLA and ALMA. The NRAO is a facility of the National Science Foundation operated under cooperative agreement by Associated Universities, Inc.

This research made use of data products from the Midcourse Space Experiment. Processing of the data was funded by the Ballistic Missile Defense Organization with additional support from NASA Office of Space Science. This research has also made use of the NASA/ IPAC Infrared Science Archive, which is operated by the Jet Propulsion Laboratory, California Institute of Technology, under contract with the National Aeronautics and Space Administration.

This publication makes use of data products from the Two Micron All Sky Survey, which is a joint project of the University of Massachusetts and the Infrared Processing and Analysis Center/California Institute of Technology, funded by the National Aeronautics and Space Administration and the National Science Foundation.




\end{document}